\documentstyle[psfig,prd,aps]{revtex}

\def\sint{\int \!\!\!\!\!\!\!\! \sum_{p}}

\def\tr{\mbox{tr}\,}
\def\Tr{\mbox{Tr}\,}

\def\be{\begin{equation}}
\def\ee{\end{equation}}
\def\bea{\begin{eqnarray}}
\def\eea{\end{eqnarray}}

\begin{document}
\draft
\preprint{TAN-FNT-98-03}

\title{Chiral quark models with non-local separable interactions
at finite temperature and chemical potential}

\author{D. G\'omez Dumm$^a$ $^\dagger$ and Norberto N. Scoccola$^{b,c}$
        \footnote[2]{Fellow of CONICET, Argentina.}}

\address{
$^a$ IFLP, Depto.\ de F\'{\i}sica, Universidad Nacional de La Plata, \\
     C.C. 67, (1900) La Plata, Argentina.\\
$^b$ Physics Department, Comisi\'on Nacional de Energ\'{\i}a At\'omica, \\
     Av.Libertador 8250, (1429) Buenos Aires, Argentina.\\
$^c$ Universidad Favaloro, Sol{\'\i}s 453, (1078) Buenos Aires, Argentina.}

\date{July 2001}

\maketitle

\begin{abstract}
\noindent Chiral quark models with non-local covariant separable
interactions at finite temperature and chemical potential are investigated.
We develop a formalism in which the different quark properties are evaluated
taking into account the analytic structure of the quark propagator. In this
framework we
study the chiral restoration phase transition for several definite non-local
regulators, including that arising within the instanton liquid picture. We
find that in all cases the chiral transition is of first order for low values
of $T$, turning into a smooth crossover at a certain ``end point''. Using
model parameters which lead to the physical pion mass and
decay constant, we find for the position of this ``end point''
the values $(T_E, \mu_E) \approx (60-70, 180-210)$~MeV. We also discuss
the special relevance of the first poles of the quark propagator.
\end{abstract}

\pacs{12.39.Ki, 11.30.Rd, 11.10.Wx}

\section{Introduction}
The understanding of the behaviour of strongly interacting matter at finite
temperature and/or density is of fundamental interest and has important
applications in cosmology, in the astrophysics of neutron stars and in the
physics of relativistic heavy ion collisions. From the theory of the
quark-gluon interactions, Quantum Chromo Dynamics (QCD), we believe that at
zero temperature and density chiral symmetry is spontaneously broken and
quarks are confined within hadrons. However, since QCD is asymptotically
free, when either the temperature $T$ or the chemical potential $\mu$ are
high the effective coupling becomes small. Thus, one expects that at a
certain point the system undergoes a phase transition (or crossover) to a
new phase in which color is screened rather than confined and chiral
symmetry is restored. Unfortunately, so far, it has not been possible to
obtain detailed information about the corresponding $T-\mu$ phase diagram
directly from QCD. In fact, lattice simulations, which work well for zero
density and finite temperature, have serious difficulties in dealing with
the complex fermion determinant that arises at finite chemical
potential\cite{Kar01}. In this situation, different models have been used to
study this sort of problems. Among them the Nambu-Jona-Lasinio
model~\cite{NJL61} is one of the most popular. In this model the quark fields
interact via local four point vertices which are subject to chiral symmetry.
If such interaction is strong enough chiral symmetry is spontaneously broken
and pseudoscalar Goldstone bosons appear. It has been shown by many authors
that when the temperature and/or density increase, the chiral symmetry is
restored\cite{VW91}. As an improvement on the local NJL model, some
covariant non-local extensions have been studied in the last few
years\cite{Rip97}. Nonlocality arises naturally in the context of several of
the most successful approaches to low-energy quark dynamics as, for example,
the instanton liquid model\cite{SS98} and the Schwinger-Dyson resummation
techniques\cite{RW94}. It has been also argued that non-local covariant
extensions of the NJL model have several advantages over the local scheme.
Indeed, non-local interactions regularize the model in such a way that
anomalies are preserved\cite{AS99} and charges properly quantized, the
effective interaction is finite to all orders in the loop expansion and
therefore there is not need to introduce extra cut-offs, soft regulators
such as Gaussian functions lead to small NLO corrections\cite{Rip00}, etc.
In addition, it has been shown\cite{BB95} that a proper choice of the
non-local regulator and the model parameters can lead to some form of quark
confinement, in the sense that the effective quark propagator has no poles
at real energies. In a previous letter\cite{GDS01} we have studied the phase
diagram of a non-local model with a Gaussian regulator. The aim of the
present work is to present details of such analysis as well as to extend it
to more general cases, such as the Lorentzian regulator and the regulator
arising within the instanton liquid model.

This article is organized as follows. In Sec.\ II we introduce the model and
the formalism needed to extend it to finite temperature and density. In
Sec.\ III we discuss how to perform the summation over the Matsubara modes
for the expressions obtained previously. Our results for some specific
non-local regulators are presented in Sec.\ IV. Finally, in Sec.\ V we give
our conclusions.

\section{Formulation}

Our starting point is the partition function at zero $T$ and $\mu$,
\begin{equation}
{\cal Z}_0 = \int {\cal D} \bar\psi {\cal D} \psi \  e^{-S_E}\;,
\label{zcero}
\end{equation}
where $S_E$ stands for the Euclidean action
\begin{equation}
S_E = \int d^4 x \left[ \bar \psi (x) \left(- i \rlap/\partial  + m_c \right)
\psi (x) - \frac{G}{2}\, j_a(x) j_a(x) \right] \,.
\end{equation}
Here $m_c$ is the current quark mass, and the euclidean operator $\rlap/\partial$
is defined as
\begin{equation}
\rlap/\partial = \gamma_4 \frac{\partial\ }{\partial\tau} + \vec\gamma \cdot
\vec\nabla
\end{equation}
with $\gamma_4 = i\gamma_0$, $\tau = i t$. The current $j_a(x)$ is given by
\begin{equation}
j_a (x) = \int d^4 y\ d^4 z \ r(y-x) \ r(x-z) \  \bar \psi(y) \Gamma_a
\psi(z)\,,
\end{equation}
where $\Gamma_a = ( 1, i \gamma_5 \vec \tau )$ and $r(x-y)$ is a non-local
regulator function. The regulator is local in momentum space, namely
\begin{equation}
r(x-z) = \int \frac{d^4p}{(2\pi)^4} \ e^{-i(x-z) p} \ r(p) \;.
\end{equation}
In fact, Lorentz invariance implies that $r(p)$ can only be a
function of $p^2$, hence we will use for the Fourier transform of
the regulator the form $r(p^2)$ from now on.

To proceed it is convenient to deal with bosonic degrees of freedom. Let us perform a standard bosonization of the
theory, introducing the sigma and pion meson fields
\begin{equation}
M_a(x) = ( \sigma(x), \vec \pi(x)) \;.
\end{equation}
Following the usual steps we obtain a partition function
equivalent to that in (\ref{zcero}):
\begin{equation}
{\cal Z} = \int {\cal D} \sigma {\cal D} \pi \  \det A(M_a)\
\exp \left[-\frac{1}{2 G} \int \frac{d^4 p}{(2\pi)^4} \ M_a^2(p) \right]\,,
\label{spi}
\end{equation}
where the operator $A$ in momentum space reads
\begin{equation}
A(M_a) = (\,-\rlap/p + m_c)\,(2\pi)^4 \,\delta^{(4)}(p-p')
+ r(p^2)\, M_a(p-p')\,r({p'}^2)\, \Gamma_a\;.
\end{equation}
At this stage, we perform the mean field approximation by expanding the meson
fields around their translational invariant vacuum expectation values
\begin{eqnarray}
\sigma(x) &=& \bar \sigma +\  \delta \sigma (x) \\
\pi_i(x) &=& \bar \pi_i +\  \delta \pi_i (x)
\end{eqnarray}
and neglecting the fluctuations $\delta \sigma (x)$ and $\delta \pi_i (x)$.
The mean values of the pion fields vanish due to symmetry reasons.
Within this approximation the determinant in (\ref{spi}) is
formally given by
\begin{equation}
\det A = \exp \Tr \log A = \exp V^{(4)} \int \frac{d^4 p}{(2\pi)^4}\;
\tr \log \left[\,-\rlap/p + m_c + \bar \sigma \, r^2(p^2)\,\right]
 \,,
\label{det}
\end{equation}
where $\tr$ stands for the trace over the Dirac, flavor and color indices,
and $V^{(4)}$ is the four-dimensional volume of the functional integral.

Now the corresponding partition function in the grand canonical ensemble
for finite temperature $T$ and chemical potential $\mu$ can be obtained
through the replacement in the integrals in (\ref{spi}) and (\ref{det})
\begin{equation}
\int \frac{d^4 p}{(2\pi)^4}\; F(p_4,\vec p) \quad \to \quad
\sint F(p_4,\vec p) \equiv
T \sum_{n=-\infty}^{n=\infty} \int \frac{d^3 p}{(2\pi)^3}\;
F(\omega_n - i\mu,\vec p)\;,
\end{equation}
where $\omega_n$ are the Matsubara frequencies corresponding to fermionic
modes, $\omega_n = (2 n+1) \pi T$. In performing this replacement we have
assumed that the regulator acquires an explicit $\mu$ dependence, as it is
the case e.g.\ in the instanton liquid model \cite{Abr81,CD99}. In the same
way we replace the volume $V^{(4)}$ by $V/T$, $V$ being the
three-dimensional volume in coordinate space. The mean field partition
function reads then
\begin{equation}
{\cal Z}_{MF}(T,\mu) = \exp \left[-\Gamma_{MF}(T,\mu) \right],
\end{equation}
where the mean field action is given by
\begin{equation}
\frac{T}{V}\,\Gamma_{MF} (T,\mu) =
- 4 N_c \left( \sint \log \left[ p^2 + \Sigma^2(p^2) \right] \right)
+ \frac{\bar \sigma^2}{2 G}
\label{action}
\end{equation}
with the quark selfenergy $\Sigma(p^2)$ defined as
\begin{equation}
\Sigma(p^2) = m_c + \bar \sigma \ r^2(p^2)\, .
\end{equation}
Finally, the grand canonical thermodynamic potential is given by
\begin{equation}
\Omega_{MF}(T,\mu) = - \frac{T}{V} \, \log {\cal Z}_{MF}(T,\mu) =
\frac{T}{V}\,\Gamma_{MF}(T,\mu)\,.
\end{equation}
It can be shown that in general this quantity turns out to be divergent.
However, it can be regularized by subtracting the corresponding value
at zero $T$ and $\mu$,
$\Omega_{MF}^{(r)} = \Omega_{MF} - \Omega_{MF}^{(0)}$. Now the minima
of the thermodynamic potential are obtained from the solutions of
\begin{equation}
\frac{\partial \Omega_{MF}}{\partial \bar \sigma} = 0
\end{equation}
which leads to the following gap equation for $\bar \sigma$:
\begin{equation}
\bar \sigma = 8 N_c G \sint \ \frac{\Sigma(p^2) \ r^2(p^2)}{p^2 +
\Sigma^2(p^2)}\;.
\label{gapequation}
\end{equation}

Given the thermodynamic potential the expressions for all other relevant quantities
can be easily derived. For each flavor the quark vacuum expectation value is given by
\begin{equation}
\langle \bar q q \rangle = \frac{ \partial \Omega_{MF}}{\partial m_c} =
- 4 N_c \sint \ \frac{\Sigma(p^2)}{p^2 + \Sigma^2(p^2)}\;,
\label{qbarq}
\end{equation}
while the quark density turns out to be
\begin{equation}
\rho_q = - \frac{ \partial \Omega_{MF}}{\partial \mu}
= -4 i N_c \ \sint \
 \frac{ p_4 + \Sigma(p^2)\ \partial_{p_4} \Sigma(p^2)}{p^2 +
 \Sigma^2(p^2)}\;.
\label{qq}
\end{equation}
In obtaining these last two equations it should be noticed that for each quark flavor
only half of the kinetic term in Eq.(\ref{action}) contributes to the derivatives.

As well known, if the quark selfenergy is momentum independent as in the
conventional NJL model, the sums over the Matsubara frequencies indicated in
the previous equations can be easily carried out and one obtains rather
simple expressions in terms of the usual occupation numbers. This also holds
when the selfenergy only depends on the spatial components of the momentum.
However, for the covariant regulators we are considering here the situation
is more complicated. The main difficulty is that the analytic structure of
the quark propagator $S(q)= 1/(- \rlap/ q + \Sigma(q^2) )$ in the complex
plane can be much richer in this case: there might be a rather complicate
structure of poles and cuts.

In Minkowski space and for $\vec q=0$ the positions of the poles are given by
the solutions of
\begin{equation}
q_0^{\ 2} - \left[ m_c + \bar \sigma \ r^2(-q_0^{\ 2}) \right]^2 = 0.
\end{equation}
Let us denote the real and imaginary parts of these solutions with $r_p$ and
$i_p$ respectively. For nonzero spatial momentum $\vec q$ it is easy to see that
the poles are located at $z_p = R_p + i I_p$, where
\begin{equation}
R_p = \pm \epsilon_p \ ,  \ \ \ \ \ \ I_p = \pm \frac{r_p i_p}{\epsilon_p}
\label{trf}
\end{equation}
with
\begin{equation}
\epsilon_p = \sqrt{ \frac{ r_p^2 - i_p^2 + \vec q^{\ 2} +
\sqrt{ (r_p^2 - i_p^2 + \vec q^{\ 2})^2 + 4\ r_p^2 \ i_p^2
} }{2}} \;.
\end{equation}
On the other hand, the possible cuts of the propagator will be given in
general by the cuts of the regulator as a function of $q$. As in the case of
the poles, it is convenient to determine first their position in the $q_0$
plane for $\vec q=0$. Then it is simple to find the position in the general
case applying the above equations to all the points along the cuts.

Before going into the explicit evaluation of the Matsubara sums it is useful
to analyze the pole and cut configurations in some relevant situations. One
important case is that in which the regulator is such that the Minkowski
quark propagator has an arbitrary set of poles but no cuts. Within this
class of regulators it might exist a situation in which there are no poles
along the real axis. As already mentioned, this situation might be interpreted
as a realization of confinement\cite{BB95}. In that case only quartets of
poles located at $\alpha_p = r_p \pm i \ i_p$, $\alpha_p = - r_p \pm i \
i_p$ appear. On the other hand, if purely real poles do exist they will show
up as doublets $\alpha_p = \pm \ r_p$. It is clear that the number and
position of the poles depend on the details of the regulator. For example,
if we assume it to be a step function as in the standard NJL model, only two
purely real poles at $\pm \ M$ appear, with $M$ being the dynamical quark
mass. For a Gaussian interaction, three different situations might occur.
These can be classified according to the value of $\Sigma(0)$ at zero $T$
and $\mu$, which we denote by $\bar\Sigma(0)$. For values of $\bar\Sigma(0)$
below a certain critical value $\Sigma(0)_{crit}$, two pairs of purely real
simple poles and an infinite set of quartets of complex simple poles appear.
At $\bar\Sigma(0) = \Sigma(0)_{crit}$, the two pairs of purely imaginary
simple poles turn into a doublet of double poles with $i_p=0$, while for
$\bar\Sigma(0) > \Sigma(0)_{crit}$ only an infinite set of quartets of
complex simple poles is obtained. In the case of the Lorentzian
interactions, there is also a critical value above which purely real poles
at low $q_0^2$ cease to exist. However, for this family of regulators the
total number of poles is always finite. As we see, several physically
interesting regulators indeed belong to the ``no cuts'' group. Unfortunately
some other important ones, like that arising within the instanton liquid
model, do lead to propagators which present poles and cuts. Thus, it is
necessary to consider this more general situation.

\section{Evaluation of Matsubara sums}

To proceed with the evaluation of the sums over the Matsubara frequencies we
assume that the regulator is such that the quark propagator at $\vec q=0$
has the analytic structure shown in Fig.~1a. Whereas the proposed pole
distribution is completely general, we consider for simplicity the
particular case in which there is a single cut lying along the real axis in
the regions $[-\infty, -r_c]$ and  $[r_c,\infty]$. Let us consider the sum
appearing in the gap equation Eq.(\ref{gapequation}),
\begin{equation}
S  \equiv \int \frac{d^3q}{(2\pi)^3}\ \ T \sum_{n=-\infty}^{\infty} \
F[(\omega_n-i\mu)^2+\vec q^{\ 2})] = \int \frac{d^3q}{(2\pi)^3}\
I(\vec q^{\ 2}) \ \ , \label{gapeq}
\end{equation}
where
\begin{equation}
F(q^2) = \frac{ \Sigma (q^2) r^2(q^2)}{q^2 + \Sigma^2(q^2)}\ .
\end{equation}
In order to evaluate the sum in (\ref{gapeq}) it is convenient to fix $\vec
q$ and define ${\cal F}(z)\equiv F[(-i z-i\mu)^2+\vec q^{\ 2})]$, so that
the argument of the sum is given by ${\cal F}(i\omega_n)$. It is easy to see
that the structure of cuts and poles of ${\cal F}(z)$ is similar to that
shown in Fig.~1a, conveniently translated to the $z$ plane (see Fig.~1b).
This translation can be performed by means of Eqs.(\ref{trf}), with an
additional shift $R_p\to R_p-\mu$. We introduce now the auxiliary function
\begin{equation}
f(z) = \left[ 1 + \exp(z/T) \right]^{-1}\;,
\end{equation}
which has a set of poles at $z_n = i \omega_n$ with $-T$ as the corresponding
residues. Provided that ${\cal F}(z)$ has no poles on the imaginary axis,
the sum we want to evaluate can be regarded as the sum over the $z_n$ poles
of the residues of the function $-{\cal F}(z) f(z)$. We can use Cauchy's theorem
to get
\begin{equation}
I(\vec q^{\ 2})  =  T \sum_{n=-\infty}^{\infty} \ {\cal F}(i \omega_n) =
\frac{i}{2\pi} \int_{-i \infty + \delta}^{i \infty + \delta} dz \ {\cal
F}(z) \ f(z) - \frac{i}{2\pi} \int_{-i \infty - \delta}^{i \infty - \delta}
dz \ {\cal F}(z) \ f(z) \;. \label{int1}
\end{equation}
The integrals in Eq.(\ref{int1}) can be calculated with the help of
complex integrals along adequate paths in the $z$ plane. Let us
consider the sum
\be
\int_{C_1} dz\ {\cal F}(z)\, f(z)\, + \,
\int_{C_2} dz\ {\cal F}(z)\, f(z)\, + \,
\int_{C_3} dz\ {\cal F}(z)\, f(z)\, \;,
\ee
where the contours $C_1$, $C_2$ and $C_3$ are shown in Fig.~1b.
In the limit $R\to\infty$, one can make use of Cauchy's theorem to obtain
\bea
I(\vec q^{\ 2}) & = & \sum_{R_p > - \mu}
\mbox{Res}\,[{\cal F}(z);z_p] \ f(z_p) \nonumber \\
& & -\, \frac{1}{\pi} \int_{\sqrt{r_c^2 + \vec q^{\ 2}} - \mu}^\infty dx
\ \Delta{\cal F}(x) \, f(x)
- \frac{i}{2\pi} \int_{-i\infty-\mu}^{i\infty-\mu}  dz\ {\cal F}(z)\,f(z)\;,
\label{caminos}
\eea
where $z_p=R_p+i I_p$ are the poles of ${\cal F}(z)$, and we have defined
\begin{equation}
\Delta{\cal F} \equiv \frac{ {\cal F}(x+i\epsilon)-
{\cal F}(x-i\epsilon)}{2 i}\;,
\end{equation}
which is different from zero along the cut. Notice that for large $R$ the
integrals over the round
contours are expected to vanish in view of the exponential fall of $f(z)$.

\vspace*{.2cm}
\begin{figure}[hbt]
\centerline{\psfig{figure=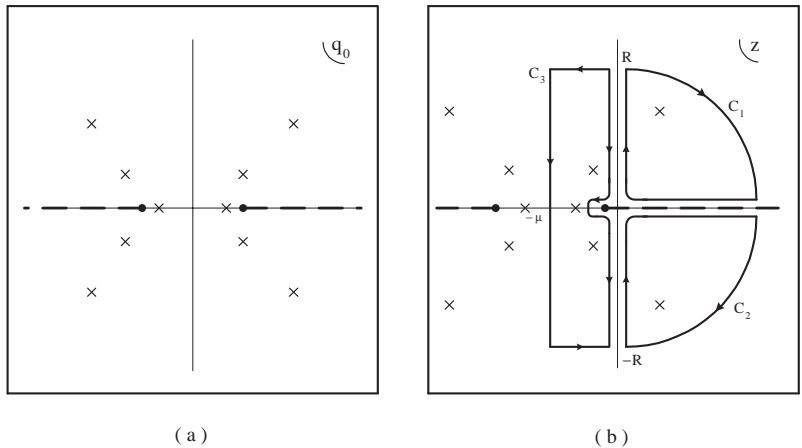,height=6cm}}
\vspace*{.4cm}
\protect\caption{\it (a) Poles and cuts of the quark propagator in the $q_0$
plane for $\vec q=0$. (b) Pole and cut structure of the propagator
in the $z$ plane and useful integration paths to evaluate the Matsubara sums.}
\label{qzplane}
\end{figure}

Now the last integral in the right hand side of (\ref{caminos}) can
be rewritten by considering the function
\be
I_0(\vec q^{\ 2}) = -\frac{i}{2\pi}
\int_{-i\infty-\mu}^{i\infty-\mu}\,dz\ {\cal F}(z)
= \frac{1}{2\pi}\int_{-\infty}^{\infty}\,dq_4\ F(q^2)\;,
\ee
which is nothing but $I(\vec q^{\ 2})$ in the limit of vanishing $T$ and $\mu$.
Notice that
\be
- \frac{i}{2\pi} \int_{-i\infty-\mu}^{i\infty-\mu}\,dz\ {\cal F}(z)\,f(z)
= I_0(\vec q^{\ 2}) + \frac{i}{2\pi} \int_{-i\infty-\mu}^{i\infty-\mu}\,dz\
{\cal F}(z)\,f(-z)\,.
\ee
The last integral in this expression can be evaluated by
closing a contour on the left side of the complex plane and
using once again Cauchy's theorem. Finally, assuming that
the regulator is an analytic function of $q^2$, we can use the
properties
\bea
\Delta {\cal F}(- x - \mu) & = & - \Delta {\cal F}(x - \mu) =
- \mbox{Im\ } {\cal F}(x - \mu + i \epsilon)
\nonumber\\
\mbox{Res}\, [{\cal F}(z);z_p] & = &
- \mbox{Res}\, [{\cal F}(z);-z_p-2 \mu] \ ,
\eea
to end up with our main result
\bea
S & = & \int \frac{d^4q}{(2\pi)^4}\ F(q^2)\
 +\ 2 \int \frac{d^3q}{(2\pi)^3}\ \mbox{Re}\!\! \sum_{\scriptsize
\begin{array}{cc}
R_p > - \mu \\
I_p \geq 0
\end{array}} \!\! \gamma_p \
\mbox{Res}\,[{\cal F}(z);z_p]\,
\left[\, n_+(z_p+\mu) + n_-(z_p+\mu) \, \right]
\nonumber \\
& & -\, \frac{1}{\pi} \int \frac{d^3q}{(2\pi)^3}\
\int_{\sqrt{r_c^2 + \vec q^{\ 2}}}^\infty\,dx\ \mbox{Im\ }{\cal F}
( x- \mu+i\epsilon)\,
\left[\, n_+(x) + n_-(x) \, \right]\,,
\label{main}
\eea
where the first term corresponds to the value of
$S$ at zero temperature an chemical potential. The coefficient
$\gamma_p$ is defined as $\gamma_p=1/2$ for $I_p=0$ and $\gamma_p=1$
otherwise, and $n_\pm(z)$ stand for the occupation number functions
\be
n_\pm(z) = f(z\mp \mu) = \frac{1}{1+\exp \left[(z \mp \mu)/T\right]}\;.
\ee
It is clear that the steps leading to Eq.(\ref{main}) can be also followed
to evaluate the sums appearing in e.g.\ the quark vacuum expectation
value and the quark density, Eqs.(\ref{qbarq}) and (\ref{qq})
respectively, once the function $F(q^2)$ is properly redefined.
In the case of the regularized thermodynamic
potential $\Omega_{MF}^{(r)}$ the situation is somewhat more complicated,
since the argument of the sum includes a logarithm that introduces cuts
in the $z$ plane outside the real axis. Nevertheless a
similar relation can be derived to calculate the corresponding Matsubara sum.

{}From Eq.(\ref{main}), it is now easy to obtain the well-known results for
the standard NJL model with a three-momentum cut-off $\Lambda$\cite{VW91}.
In fact, at the level of mean field approximation such model can be viewed
as a particular case of the non-local schemes analyzed above, where the
regulator is taken as a theta function in three-momentum space, $r(q^2) =
\theta(\vec q^{\ 2} -\Lambda^2)$ ({\em  i.e.} it is not covariant). Notice
that for this regulator there is only one pole at $z_0 =
\sqrt{\vec q^{\ 2}+M^2}-\mu$, $M=m_c+\bar\sigma$, with Res$[{\cal F}(z);z_0] =
-M/2(z_0+\mu)$. From the relation $S = \bar \sigma/(8GN_c)$ (see
Eqs.(\ref{gapeq}) and (\ref{gapequation})) one immediately arrives at the
standard NJL gap equation
\begin{equation}
M = m_c + \frac{2 G N_c}{\pi^2}\int_0^\Lambda dp\; p^2\; \frac{M}{E_p}
\left[ 1 - n_+(E_p) - n_-(E_p)\right]\,,
\end{equation}
where $E_p=\sqrt{p^2+M^2}$.

\section{Results for definite non-local regulators}

Having introduced the formalism needed to deal with models with non-local
separable interactions at finite temperature and chemical potential,
we turn now to show the results of numerical calculations carried
out for some particular regulators. We concentrate here in three cases:
the Gaussian regulator, the Lorentzian regulator and the regulator
arising within the instanton liquid model.

\subsection{Gaussian regulator}

Let us analyze a regulator function of the form
\begin{equation}
r(q^2) = \exp\left( - q^2/2\Lambda^2 \right)\,,
\end{equation}
where $\Lambda$ is a parameter of the model, besides the coupling constant
$G$ and the current quark mass $m_c$. We consider here two sets of values
for the parameters. Set I corresponds to $G~=~50$~GeV$^{-2}$,
$m_c~=~10.5$~MeV and $\Lambda~=~627$~MeV, while for Set II the respective
values are $G~=~30$~GeV$^{-2}$, $m_c~=~7.7$~MeV and $\Lambda~=~760$~MeV.
Both sets of parameters lead to the physical values of the pion mass and
decay constant. At zero temperature and chemical potential, the calculated
values of the chiral quark condensate are $- (200$~MeV$)^3$ for Set I and $-
(220$~MeV$)^3$ for Set II. These values are similar in size to those
determined from lattice gauge theory or QCD sum rules. The corresponding
results for the selfenergy at $T=\mu=0$ and zero momentum are $\bar\Sigma
(0) = 350$~MeV for Set I and $\bar\Sigma (0) = 300$~MeV for Set II.

As stated in Sect.~II, in the case of a Gaussian regulator the quark
propagator has no purely real poles if $\bar\Sigma(0)$ is above a critical
value $\Sigma(0)_{crit}$, whereas for $\bar\Sigma(0) < \Sigma(0)_{crit}$ two
pairs of real poles exist. From the explicit expression for the critical
selfenergy,
\begin{equation}
\Sigma(0)_{crit} = m_c + \frac{1}{2} \left( \sqrt{m_c^2 + 2 \Lambda^2} - m_c
\right) \exp\left[ - \left( \sqrt{m_c^2 + 2 \Lambda^2} + m_c
\right)^2/4\,\Lambda^2\right]\,, \label{crit}
\end{equation}
it can be easily checked that Set I and Set II correspond to the first and
second situations, respectively. Thus, following Ref.\cite{BB95}, Set I
might be interpreted as a confining one, since quarks cannot materialize
on-shell in Minkowski space.

\begin{figure}[hbt]
\centerline{\psfig{figure=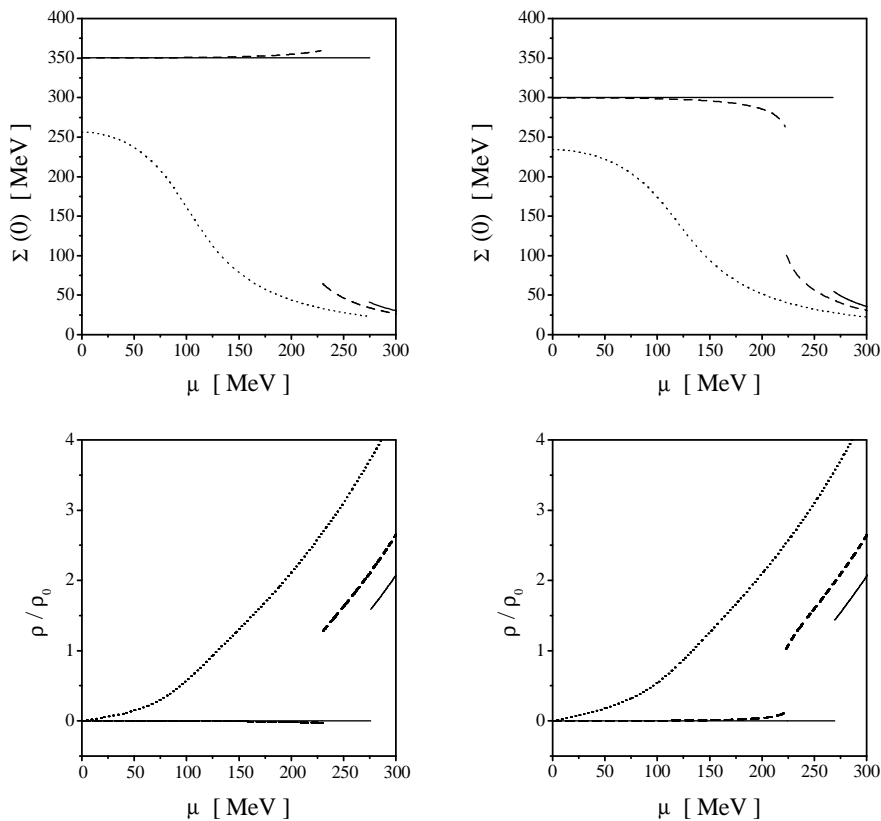,height=11cm}}
\protect\caption{\it Behaviour of the selfenergy (upper graphs) and the
density (lower graphs) with the chemical potential for three representative
values of the temperature. Full lines corresponds to $T=0$, dashed lines to
$T=50$ {\rm MeV} and dotted lines to $T=100$ {\rm MeV}. The left panels
display the results for Set I and the right panels those for Set II.}
\label{tmdep}
\end{figure}

The behaviour of the zero-momentum selfenergy $\Sigma(0)$ as a function of
the temperature and chemical potential can be obtained by solving the gap
equation for finite $T$ and $\mu$, Eq.(\ref{gapequation}), where the
Matsubara sum can be evaluated using Eq.(\ref{main}). In this case the
regulator has no cuts in the complex plane, hence the last term of
(\ref{main}) vanishes. A similar procedure can be followed to obtain the
quark density $\rho$ for finite $T$ and $\mu$ from Eq.(\ref{qq}). Our
numerical results are shown in Fig.~2, where we plot $\Sigma(0)$ vs.\ $\mu$
and $\rho$ vs.\ $\mu$ for some fixed values of the temperature. The values
of $\rho$ are given with respect to nuclear matter density, $\rho_0 \simeq
1.3\times 10^6$~MeV$^3$. The left and right panels in the figure correspond
to the results for Set I and Set II, respectively. For both sets of
parameters we observe the existence of some kind of phase transition at (or
around) a given value of the chemical potential which depends on the
temperature. The only qualitative differences appear in the behaviour of
$\Sigma(0)$ and $\rho$ as functions of $\mu$ for low temperatures. For
example, for $T=50$~MeV we observe that in the case of the confining set
(Set I) the selfenergy rises with $\mu$ (for $\mu$ below the critical
chemical potential) while the opposite occurs in the case of Set II. As one
increases the temperature the behaviour of Set I turns into that of Set II
and, eventually, in both cases we obtain a continuous and always decreasing
curve. An equivalent behaviour is observed in the case of the density. To
obtain our results, we have included in the sums over the poles of the quark
propagator only the first few poles close to the origin in the $q_0$ plane.
We have checked, however, that for the range of values of $T$ and $\mu$
covered in our calculations, the convergence is so fast that already the
first pole gives almost 100\% of the full result. Thus, the behaviour of the
relevant physical quantities up to (and somewhat above) the phase transition
is basically dominated by the first set of poles (doublet or quartets) of
the propagator.

We observe that at $T=0$ there is a first order phase transition for both
the confining and the non-confining sets of parameters. As the temperature
increases, the value of the chemical potential at which the transition shows
up decreases. Finally, above a certain value of the temperature the first
order phase transition does not longer exist and, instead, there is a smooth
crossover. This phenomenon is clearly shown in the left panel of Fig.~3,
where we display the critical temperature at which the phase transition
occurs as a function of the chemical potential. The point at which the first
order phase transition ceases to exist is usually called ``end point''.

\begin{figure}[hbt]
\centerline{\psfig{figure=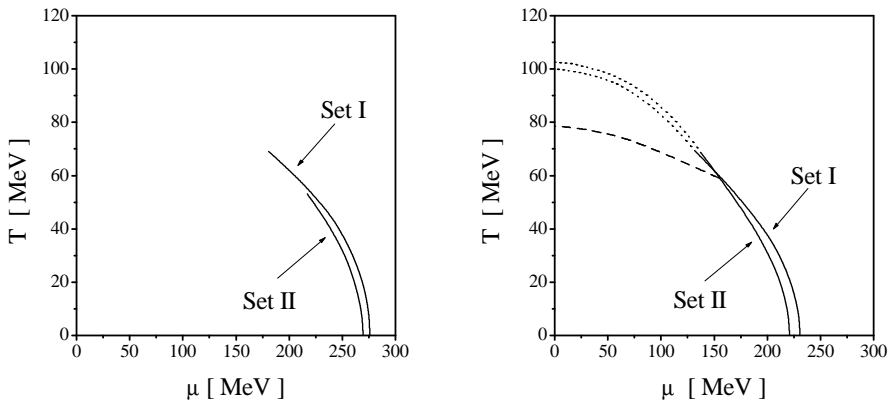,height=5.5cm}} \protect\caption{\it
Critical temperatures as a function of the chemical potential. The left
panel corresponds to the case of finite quark masses and the right panel to
the chiral limit. In the right panel, the dotted lines correspond to the
region of the second order phase transition, while the full lines in both
panels indicate the region where a first order transition occurs. The dashed
line in the right panel indicates the critical line for deconfinement
corresponding to Set I. For chemical potentials somewhat larger than $\mu_P$
this line coincides with that of the chiral restoration.} \label{trans}
\end{figure}

In the chiral limit, the ``end point'' is expected to turn into a so-called
``tricritical point'', where the second order phase transition expected to
happen in QCD with two massless quarks becomes a first order one. Indeed,
this is what happens within the present model for $m_c=0$, as it is shown in
the right panel of Fig.~3. Some predictions about both the position of the
``tricritical point'' and its possible experimental signatures exist in the
literature\cite{BR98,HJSSV98}. In our case this point is located at $(T_P,\mu_P) =
$(70~MeV$,\,130$~MeV) for Set I and (70~MeV$,\,140$~MeV) for Set II, while
the ``end points'' are placed at $(T_E,\mu_E) = $(70~MeV$,\,180$~MeV) and
(55~MeV$,\,210$~MeV), respectively.

In general, for temperatures below the end point there is a range of values
of $\mu$ where one finds two solutions of the gap equation, {\em i.e.} there
are two values of the selfenergy where the thermodynamic potential shows a
local minimum. The situation is illustrated in Fig.~4, where we plot the
(renormalized) thermodynamic potential vs.\ the selfenergy for different
values of the temperature and some fixed values of the chemical potential
around the phase transition point. The curves in the figure correspond to
the parameters of Set~II. In the left panel we consider the case of $T=0$,
where the phase transition occurs at $\mu_c\simeq 270$ MeV, and we plot the
values of $\Omega^{(r)}$ for values of $\mu$ which are 10 MeV above and
below $\mu_c$. The equilibrium point is clearly that corresponding to the
absolute minimum in each case. In the central panel we take $T=50$ MeV,
close to the end point temperature of $\sim\,55$ MeV, hence the two minima
approach each other and the thermodynamic potential between them is
approximately constant. For the right panel the temperature is 100 MeV, well
above the end point, therefore one finds only one minimum and the phase
transition proceeds through a smooth crossover.

\begin{figure}[hbt]
\centerline{\psfig{figure=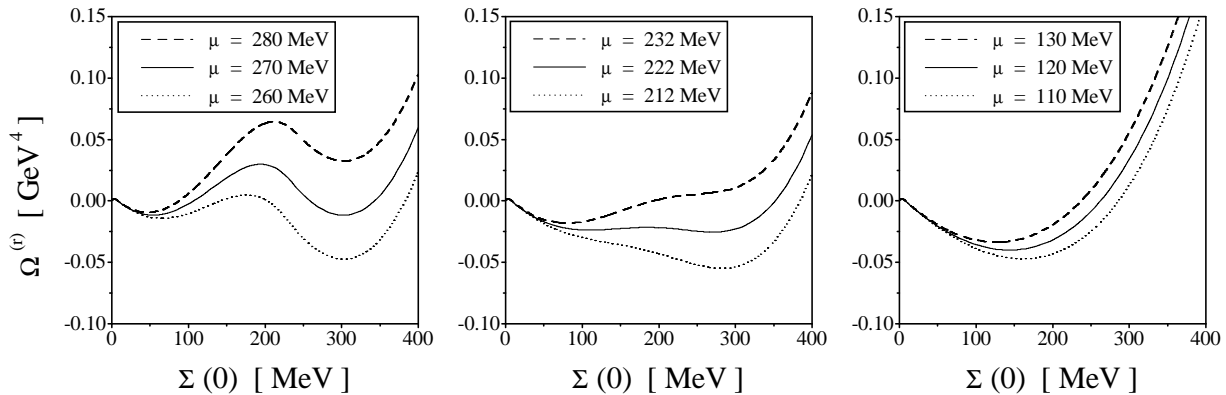,height=5.5cm}}
\protect\caption{\it Renormalized thermodynamic potential for Set II as a function
of the selfenergy, for different values of the chemical potential $\mu$. Left,
central and right panels correspond to $T=0$, $T=50$ {\rm MeV} and
$T=100$ {\rm MeV} respectively.}
\label{omega}
\end{figure}

The behaviour of the critical temperatures with the quark density is shown
in Fig.~5, where we plot the $T-\rho$ phase diagrams for both the confining
and the non-confining sets. The curves in the left panel correspond to the
case of finite quark masses, and the fat dots indicate the end points for
each set of parameters. The area below each curve is a region where both
phases are allowed. This ``mixed phase'' zone can be interpreted~\cite{BR98}
as a region where droplets containing light quarks of mass $m_c$ coexist
with a gas of constituent, massive quarks. The corresponding curves in the
chiral case ($m_c=0$) for both Sets I and II are shown in the right panel of
the figure. Here the dotted lines correspond to the second order transitions,
and the fat dots indicate the tricritical points.

\begin{figure}[hbt]
\centerline{\psfig{figure=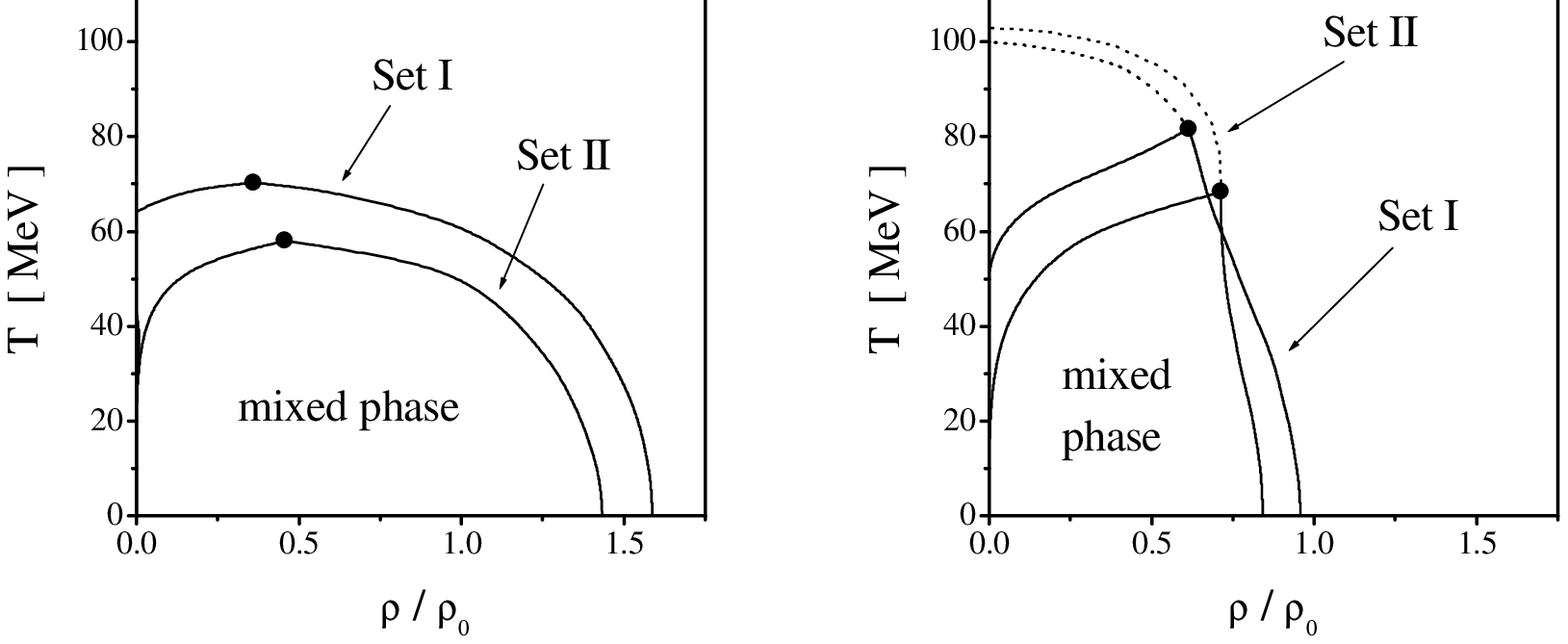,height=5.5cm}}
\protect\caption{\it Critical temperatures as functions of the relative
density for Sets I and II. The left panel corresponds to the case of finite
quark masses and the right panel to the chiral limit. The fat dots indicate
the end points (left panel) and tricritical points (right panel), and the
dotted lines in the right panel indicate the second order phase
transitions.} \label{rhogauss}
\end{figure}

It is interesting to discuss in detail the situation concerning the confining set.
In this case we can find, for each temperature, the chemical potential $\mu_{d}$
at which confinement is lost. Following the proposal of Ref.\cite{BB95}, this
corresponds to the point at which the selfenergy at zero momentum reaches the
value $\Sigma(0)_{crit}$. Using the values of $m_c$ and $\Lambda$ corresponding
to Set I we get $\Sigma(0)_{crit} = 267$~MeV. We see that for low temperatures
the value of $\mu_d$ coincides with the chemical potential corresponding to
the chiral restoration, while for temperatures close enough to that of the
``end point'' it starts to lie slightly below the phase transition line.
Above the end point temperature $T_E$ it is difficult to make
an accurate comparison since, for finite quark masses, the chiral restoration
proceeds through a smooth crossover. However, we can still study the situation
in the chiral limit. We find that in the region where the chiral phase
transition is of second order deconfinement always occurs, for fixed $T$,
at a lower value of $\mu$ than the chiral restoration. The corresponding
critical line is indicated by a dashed line in the right panel of Fig.~3. In
any case, as we can see in this figure, the departure of the line of chiral
restoration from that of deconfinement is in general not too large. This
indicates that within the present model both transitions tend to happen at,
approximately, the same point.

\subsection{Lorentzian regulator}

Let us now consider the case of a Lorentzian regulator,
\begin{equation}
r(q^2) = \frac{1}{1 + (q^2/\Lambda^2)^n}\,,
\end{equation}
where $n$ is a positive integer. For this family of regulators, it can be
seen that the quark propagator always has a finite number of poles in the
$q_0$ plane. If $n$ is
even, there are $2n$ quartets of complex poles, plus a
doublet of real poles. For odd values of $n$,
one has a doublet (quartet) of real (complex) poles
for $\bar\Sigma(0)$ below (above) a certain critical value
$\bar\Sigma(0)_{crit}$, plus a set of $2n-1$
quartets of complex poles, and a doublet of real poles. This last doublet
occurs at $q_0^2\agt\Lambda^2$.

In order to analyze the characteristics of the chiral phase transition for
this shape of the regulator, we have taken the simplest case $n=1$, choosing
as input parameter the quark selfenergy at zero temperature and chemical
potential, $\bar\Sigma(0)$. As in the case of the Gaussian regulator, the
parameters $G$, $m_c$ and $\Lambda$ have been fixed so as to reproduce
correctly the physical values of the pion mass and decay constant. For a
typical selfenergy $\bar\Sigma(0)=300$ MeV we find the reasonable values
$G=28.38$ GeV$^{-2}$, $m_c=4.57$ MeV and $\Lambda=940$~MeV. The value of
$\bar\Sigma(0)_{crit}$ is in this case approximately 270 MeV, hence the
fermion propagator has two quartets and one doublet of poles. This is
represented in Fig.~6.

\begin{figure}[hbt]
\centerline{\psfig{figure=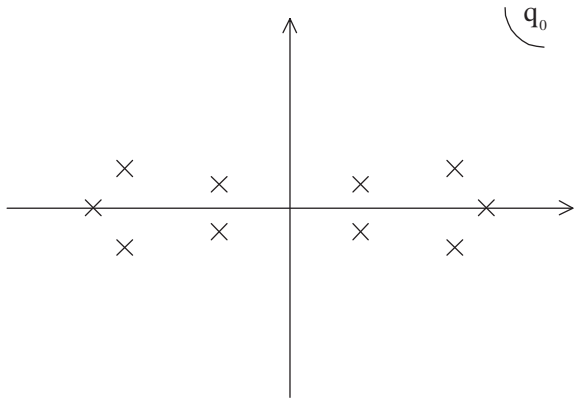,height=5.5cm}}
\vspace{0.2cm}
\protect\caption{\it Analytic structure of the quark propagator
in the $q_0$ plane for an $n=1$ Lorentzian regulator.}
\label{qlor}
\end{figure}

Once again, we can make use of both the gap equation and our result
Eq.(\ref{main}) to evaluate the quark selfenergy $\Sigma(0)$ for given
values of the temperature and chemical potential. The corresponding phase
transition curves are plotted in the left panel of Fig.~7, where we have
considered fixed temperatures of 0, 30, 50 and 100 MeV. As in the previous
cases, we observe that for low values of $T$ the chiral phase transition is
of first order, while for values above a certain $T_E$ the transition
proceeds through a smooth crossover. In the right panel of the figure we
plot for the same values of the temperature the relative quark density
$\rho/\rho_0$ as a function of the chemical potential.

\begin{figure}[hbt]
\centerline{\psfig{figure=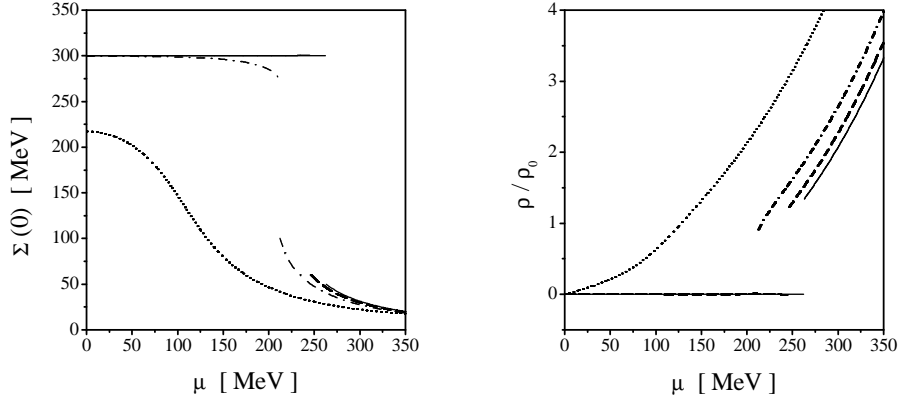,height=5.5cm}}
\protect\caption{\it
Behaviour of the quark selfenergy (left) and the relative quark density
(right) with the chemical potential in a non-local NJL model with a Lorentzian
regulator, for some representative values of $T$. Full, dashed, dashed-dotted
and dotted lines correspond to $T=0$, $30$, $50$ and $100$ {\rm MeV}
respectively.}
\label{tmdeplor}
\end{figure}

It is interesting to point out that for temperatures quite below $T_E$ (as
e.g.\ 30 MeV for the present input parameters) and $\mu < \mu_c$ the
selfenergy rises for increasing $\mu$. Although the growth is too slight to
be clearly perceived in the figure, it becomes evident at a magnified scale.
Then, as $T$ becomes closer to $T_E$, this behaviour turns into a decreasing
one (see the $T=50$~MeV curve in the left panel of Fig.~7), ending up in the
smooth crossover at $T > T_E$. As mentioned in Sec.\ IV.a for the Gaussian
interaction, the growing behaviour at low $T$ is characteristic for the
situation in which the Minkowski quark propagator has no real poles. We have
stated above that for the Lorentzian regulator there is always a real pole
close to $q_0^2 \agt\ \Lambda^2$. Thus, we can argue that the rise of
$\Sigma(0)$ is mainly related to the absence of real poles at low $q_0^2$,
which are in fact the poles to be associated with deconfined quarks.
Moreover, this can be understood as a new evidence that the basic features
of the $T-\mu$ behaviour of the quark properties are greatly determined by
the first pole of the quark propagator.

The values of the critical temperature as a function of the chemical
potential and the quark density are displayed in the left and right
panels of Fig.~8, respectively. Once again, the $T-\rho$ phase diagram shows
a mixed phase region. The predicted position of the end point is in this
case $(T_E,\mu_E) = $(58~MeV$,\,195$~MeV).

\begin{figure}[hbt]
\centerline{\psfig{figure=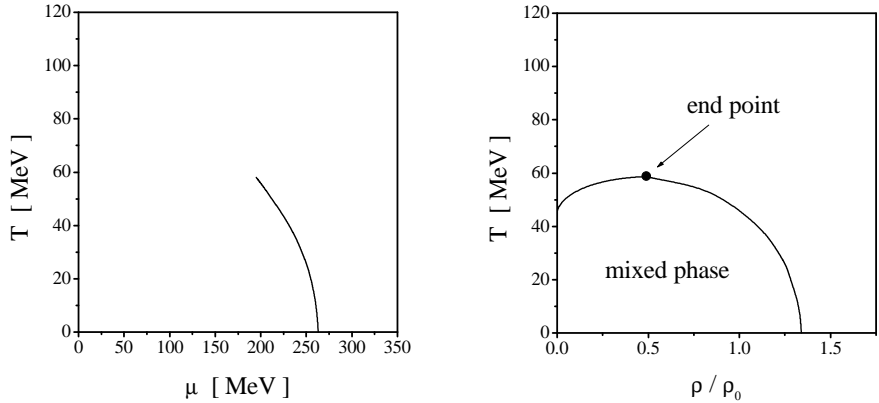,height=5.5cm}} \protect\caption{\it
Critical temperature for the Lorentzian regulator as a function of the
chemical potential (left) and the relative quark density (right).}
\label{rhodeplor}
\end{figure}

\subsection{Instanton liquid model regulator}

The instanton liquid picture of the QCD vacuum predicts a separable
non-local interaction with a regulator given by
\begin{equation}
r(q^2) = \left. - x \, \frac{d}{dx} \left[ I_0(x) K_0(x) -
I_1(x) K_1(x) \right]\right|_{x=\sqrt{q^2}\rho/2}\;,
\label{instreg}
\end{equation}
\noindent where $I$ and $K$ are the modified Bessel functions and $\rho$
stands for the average instanton size. Here we take the standard value
$\rho = 1/3$~fm, and fix the coupling constant to $G= 36.7$~GeV$^{-2}$ so
as to reproduce the typical value for the instanton density
$n\approx 1$~fm$^{-4}$\cite{SS98}. Using these values, together
with $m_c= 4.9$~MeV, it is possible to reproduce the empirical values
of the pion mass and decay
constant. The resulting value of the chiral quark condensate at zero
temperature and chemical potential is $- (256$~MeV$)^3$.

As in the previous cases, we can solve the gap equation to obtain the
phase transition curve for the quark selfenergy $\Sigma(0)$. Here the
situation is more complicated than for the Gaussian and Lorentzian
cases, since the quark propagator has now a cut in the complex plane.
As discussed in Sect.~II, it is
useful to look at the regulator function (\ref{instreg}) in Minkowski space
for zero three-momentum, $r(-q_0^2)$. If we consider the analytic
extension in the complex $q_0$ plane, we see that the regulator shows a cut
along the real axis, namely
\begin{equation}
r\left(-(x+i \epsilon)^2\right) - r\left(-(x-i\epsilon)^2\right)
= 2\, i\,\pi
\left[2\,\tilde x J_0(\tilde x) J_1(\tilde x) - J_1(\tilde x)^2\right]\,,
\end{equation}
with $\tilde x = \rho\, x/2$. It is easy to see that this cut is also present
for the full quark propagator function. As stated in Sect.~III, this
translates into a cut on the real axis in the $z$ plane if we set
$q^2 = - (z+\mu)^2+\vec q ^{\ 2}$. In addition, it can be seen that
the quark propagator has
a finite number of poles. The latter come in quartets, and their number increases
with the value of the dimensionless parameter $\lambda\equiv\rho\,\bar\Sigma(0)$.
One can find a set of critical values $\lambda_i$, such that there is only one
quartet of complex poles for $0<\lambda<\lambda_0$, two quartets for
$\lambda_0<\lambda<\lambda_1$, etc. The analytic
structure of the propagator for the case of two quartets is illustrated in
Fig.~9. The critical values $\lambda_i$ only depend
on the scaled current quark mass, $\rho\, m_c$. Taking $\rho$ and $m_c$ as above
we get for the first critical values $\lambda_0\simeq 0.81$, $\lambda_1\simeq 1.09$,
$\lambda_2\simeq 1.48$. With the chosen value $G= 36.7$~GeV$^{-2}$ the gap equation
for $T=\mu=0$ leads to $\bar\Sigma(0) = 359$ MeV, thus in our case
$\lambda\simeq 0.6$ and we only have to consider a single quartet of complex
poles. This means that the sum in Eq.(\ref{main}) contains only one term.

\begin{figure}[hbt]
\centerline{\psfig{figure=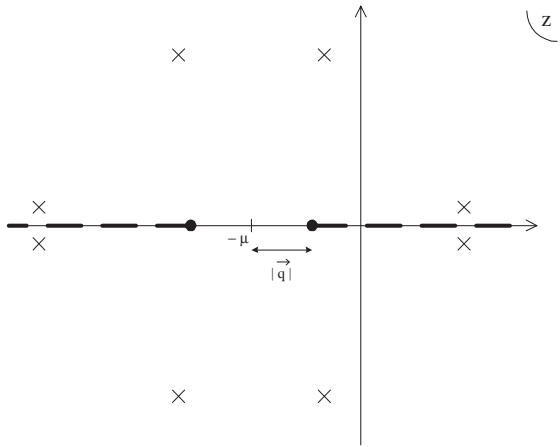,height=6cm}}
\protect\caption{\it Analytic structure of the quark propagator
for the instanton liquid model.}
\label{zins}
\end{figure}

\vspace{-.2cm}

\begin{figure}[hbt]
\centerline{\psfig{figure=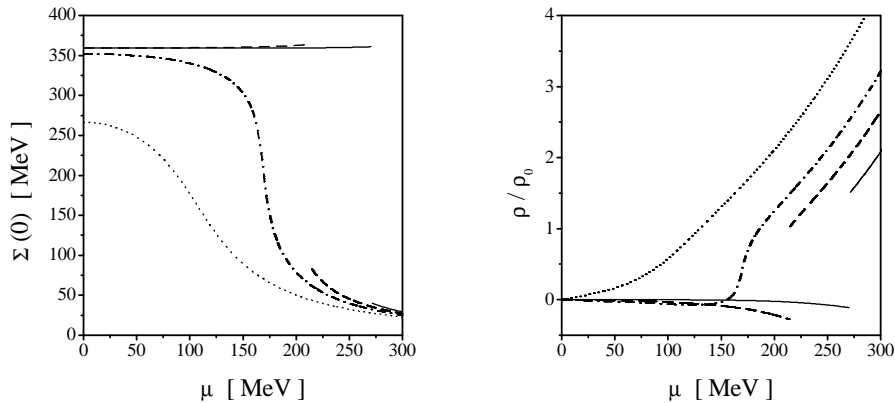,height=5.5cm}} \protect\caption{\it
Behaviour of the quark selfenergy (left) and the relative quark density
(right) with the chemical potential for the instanton liquid model. Full,
dashed, dashed-dotted and dotted lines correspond to $T=0$, $50$, $70$ and
$100$ {\rm MeV} respectively.}
\label{tmdepins}
\end{figure}

The behaviour of the quark selfenergy and the quark density as functions of
the chemical potential are shown in Fig.~10 for some representative values
of the temperature. Once again, we find that the transition is of first
order for low values of $T$, turning into a smooth crossover above a certain
end point. As in the confining Gaussian and Lorentzian cases, we find that
for finite temperatures quite below $T_E$ the selfenergy rises with $\mu$
for $\mu < \mu_c$. Thus, this feature does not seem to be dependent on the
existence of the cut, though we have checked that its contribution to the
corresponding gap equation is not negligible. Finally, as in the previous
cases, we present the curves corresponding to the phase transition
temperatures as functions of $\mu$ and $\rho/\rho_0$. These are shown in
Fig.~11. For the instanton liquid model the end point is found to be located
at $(T_E,\mu_E) = $(65~MeV$,\,180$~MeV).

\begin{figure}[hbt]
\centerline{\psfig{figure=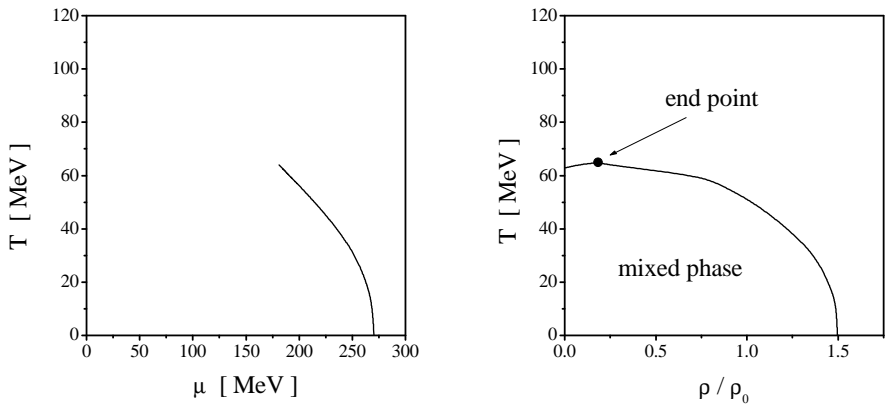,height=5.5cm}} \protect\caption{\it
Critical temperature for the instanton liquid model as a function of the
chemical potential (left) and the relative quark density (right).}
\label{rhodepins}
\end{figure}

\section{Summary and conclusions}

In this work we have studied the behaviour of chiral quark models with
non-local covariant separable interactions at finite temperature and
chemical potential. After introducing the general formalism we have derived
some expressions that allow to compute the relevant quantities in terms of
the analytic structure of the quark propagator. Although, in principle, the
sum over the Matsubara frequencies could have been directly performed in a
numerical way, we believe that our method allows for a better understanding
of some features of the chiral restoration transition (as e.g.\ the
importance of the first pole of the quark propagator) and stresses the
connection with the usual procedure followed in the standard NJL model. In
our numerical calculations we have considered three types of regulators: the
Gaussian regulator, the Lorentzian regulator and the instanton liquid model
regulator. In all these cases we have set the model parameters so as to
reproduce the empirical values of the pion mass and decay constant and to
get a chiral quark condensate in reasonable agreement to that determined
from lattice gauge theory or QCD sum rules. We find that in all cases the
phase diagram is quite similar. In particular, we obtain that for two light
flavors the transition is of first order at low values of the temperature
and becomes a smooth crossover at a certain ``end point''. Our predictions
for the position of this point are very similar for all the regulators and
slightly smaller than the values in Refs.\cite{BR98,HJSSV98},
$T_E~\approx~100$~MeV and $\mu_E \approx 200 - 230$~MeV. In this sense, we
should remark that the model under consideration predicts a critical
temperature at $\mu = 0$ of about 100~MeV\cite{BBKMT00}, somewhat below the
values obtained in modern lattice simulations which suggest $T_c \approx 140
- 190$~MeV\cite{Kar01}. In any case, our calculations seem to indicate that
$\mu_E$ might be smaller than previously expected even in the absence of
strangeness degrees of freedom.

An interesting feature of the behaviour of the quark selfenergy as a
function of the chemical potential for temperatures well below $T_E$ is that
it raises with $\mu$ (for $\mu < \mu_c)$ when the first pole in the quark
propagator is complex. A similar behaviour is found for the chiral
condensate\cite{GDS01}. This is the opposite to what happens when the first
pole is real (in Minkowski space) as in the standard NJL model\cite{VW91} or
models where the nonlocality arises only in the space components\cite{SBK94}.
Lattice calculations, even at very low chemical potential as e.g.\ those
reported in Ref.\cite{QCDTARO}, might be helpful to understand what happens
in QCD.

Several extensions of this work are of great interest. For example, it would
be very important to investigate the impact of the introduction of
strangeness degrees of freedom and flavor mixing on the main features of the
chiral phase transition. In fact, some work on an $SU(3)$ extension of the
present type of models at finite $\mu$, but $T=0$, has been recently
reported\cite{GBKG01}. Another topic that has recently attracted a lot of
attention, and is suitable to be investigated in the present model, is
the competition between chiral symmetry breaking and color superconductivity
at large chemical potential. We expect to report on these issues in future
publications.

\acknowledgements

This work was supported in part by Fundaci\'on Antorchas and Fundaci\'on
Balseiro, Argentina.

\end{document}